\documentclass[preprint,12pt]{elsarticle}
\usepackage{geometry}
\usepackage{graphicx}
\usepackage{subfigure}
\usepackage{amssymb}
\usepackage{amsmath}
\usepackage{units}
\usepackage{color}
\usepackage[nodots]{numcompress}

\journal{XX}

\begin{document}
\begin{frontmatter}


\title{Growth, microstructure and thermal transformation behaviour of epitaxial Ni-Ti films}

\author[ifw,act]{S.~Kauffmann-Weiss}
\author[TUC1]{S.~Hahn}
\author[ifw]{C.~Weigelt}
\author[ifw]{L.~Schultz}
\author[TUC1]{M.F.-X.~Wagner}
\author[ifw,TUC2]{S.~F\"ahler\corref{CorrAuthor}}
\ead{s.faehler@ifw-dresden.de}

\address[ifw]{IFW~Dresden, P.O.~Box~270116, 01171~Dresden, Germany}
\address[act]{current address: Karlsruher Institute~of~Technology, Institute for Technical Physics, 76344~Eggenstein-Leopoldshafen, Germany}
\address[TUC1]{Institute of Materials Science and Engineering, Chemnitz University of Technology, 09107~Chemnitz, Germany}
\address[TUC2]{Institute~of~Physics, Chemnitz University of Technology, 09107~Chemnitz, Germany}

\cortext[CorrAuthor]{Corresponding author}

\begin{abstract}
Epitaxial films have the potential to be used as model systems for fundamental investigations on the martensitic transformation in binary NiTi. In this paper, we discuss growth of binary NiTi thin films on single crystalline MgO substrates. Sputter deposition is used to grow NiTi films. Films prepared by complementary preparation routes (with different deposition temperatures and subsequent heat treatments) are investigated by X-ray diffraction, electron microscopy, atomic force microscopy, and electrical resistivity measurements, with the aim of optimizing film properties, particularly to obtain a well defined orientation of the austenitic unit cell and smooth surfaces. Our results show that deposition at elevated temperatures and carefully controlled subsequent heat treatments allow to produce epitaxially grown and smooth NiTi films that exhibit reversible one- or two-step martensitic transformations.
\end{abstract}


\end{frontmatter}
\newcommand{\degree}{\ensuremath{^\circ}C}
\newcommand{\td}{\ensuremath{\vartheta_{\text{D}}}}
\newcommand{\ta}{\ensuremath{\vartheta_{\text{A}}}}
\newcommand{\temp}{\ensuremath{\vartheta_{\text{T}}}}

\section{Introduction}
\label{sec:introduction}
Epitaxial films are model systems that can be used for fundamental investigations which allow for a better understanding of shape memory alloys (SMA). This idea has been extensively applied to research on magnetic shape memory alloys where epitaxial films have helped to measure the electronic origin of the martensitic transformation~\cite{Klaer-2011}, determine twin boundary energies~\cite{Anett-2011}, examine hierarchically twinned microstructures~\cite{Kaufmann-2011}, and demonstrate the adaptive nature of modulated martensite~\cite{Kaufmann-2010}. Most of these research efforts have simply been a by-product of the need for epitaxial films for microactuators~\cite{Dong-2004}, which represent the thin film counterpart of single crystals and which are required to reach the highest strains in magnetic shape memory alloys. 

In contrast NiTi, the first and most important technologically applied shape memory alloy, can be used as polycrystalline material. A good summary on polycrystalline SMA films can be found in the textbook of Miyazaki~\cite{Miya-2009}. Typically, polycrystalline films are deposited onto amorphous and/or oxidized substrates such as Si$_2$O or Si and subsequently annealed~\cite{Winzek2004,Martins2010,Kabla20104}. So far only a few studies have examined the epitaxial growth of NiTi films on heated substrates~\cite{Buschbeck2011,Martins2008}. One open research question is if epitaxial NiTi films can be used to contribute to a more fundamental understanding of the basic behaviour of NiTi SMA, as sketched above for magnetic SMA. Moreover, epitaxial NiTi films might even improve the functionality compared to polycrystalline NiTi films, which have already been developed towards microactuators~\cite{Kohl-2010}, stents for brain surgery~\cite{Miranda-2009}, elastocaloric refrigeration~\cite{Bechtold-2012}, or used for the combinatorial search for low hysteresis materials~\cite{Zarnetta-2012,Cui-2006}. 

Fundamental as well as applied research on epitaxial NiTi films will benefit from two main aspects, both of which we address in this paper: First, a well-defined orientation of the austenitic unit cell in the epitaxial films is required. This allows probing anisotropic properties and identifying orientations within the martensitic microstructure after cooling. Second, a smooth film surface is essential, as only such a well-defined cut through the crystal allows distinguishing a martensitic microstructure from film morphology later on. Small scale mechanical testing also requires smooth surfaces to avoid crack formation in tensile tests \cite{Hahn2016_1} or adverse effects on nanoindentation measurements \cite{Pfetzing2013,Pfetzing2010}. Both aspects can be examined best when the film is within the austenitic state at room temperature and therefore in the present study we focus on a film composition where the martensitic transformation occurs just below room temperature. We compare three different preparation routes: \textbf{A} depositing at room temperature followed by annealing; \textbf{B} depositing at high temperatures; \textbf{C} depositing at a moderate temperature followed by an annealing and temper step, and we investigate their effect on microstructures, surface features and thermal transformation behaviour in the resulting epitaxial NiTi films.

\section{Experimental}
\label{sec:experimental}
Physical vapour deposition based on a DC magnetron sputtering process was used with a Ni$_{46.8}$Ti$_{53.2}$ alloyed target for deposition at different temperatures. The base pressure was lower than $<\unit[10^{-8}]{mbar}$. A mixture of argon and hydrogen was used as protective atmosphere and a sputter power of \unit[100]{W} was applied. In order to induce an epitaxial growth of the deposited NiTi films we used single crystalline MgO(001) substrates. The films were grown to a thickness of \unit[120]{nm} at an average growth rate of \unit[0.16]{nm/s}.

In this study we systematically discuss results for three main preparation routes: In a first series, films were deposited at room temperature. Subsequently, the films were annealed between \unit[450]{\degree} and \unit[850]{\degree} for different times to promote thin film crystallisation. This preparation route is referred to hereafter as route \textbf{A}. In a second series (route \textbf{B}), substrate temperatures between 150 and \unit[650]{\degree} were used in order to enable epitaxial growth and chemical ordering during the deposition process itself. For the third series (route \textbf{C}), films were deposited at \unit[250]{\degree} and subsequently annealed. An additional tempering at \unit[300]{\degree} for \unit[20]{min} was applied. All annealing and tempering steps were performed in-situ in the deposition chamber with a pressure of $<\unit[10^{-6}]{mbar}$. Fast cooling rates of $r>\unit[50]{\degree/min}$ were used in order to suppress decomposition of the NiTi films. The composition of all films was determined as about Ni$_{50.1}$Ti$_{49.9}$ ($\pm  0.1\;\!at \, \%$). Comparing target and thin film compositions, we note a loss of Ti during deposition, which is a well-known effect \cite{Rumpf2004}.

The following analytical methods were used to characterise the chemical composition, microstructure and topography of the samples: For energy dispersive X-Ray spectroscopy (EDX) measurements, a Philips scanning electron microscope (SEM) with a Ni$_{50}$Ti$_{50}$ standard was used. Phase analysis was carried out using X-ray diffraction (XRD) in a Bruker D8 X-Ray diffractometer with Co-K$_\alpha$ radiation. To analyse global textures, pole figures were measured using a four circle set-up (Philips X'Pert) in the range of $0^\circ\leq\varphi\leq\,360^\circ$ and $\psi\leq\,80^\circ$ in steps of $1^\circ$ with Cu-K$_\alpha$ radiation. Based on the pole figure measurement results, complementary local texture measurements were performed by electron backscatter diffraction (EBSD) with a NEON40EsB (Zeiss) SEM. Sample surfaces were examined using a DI Dimension 3100 atomic force microscope (AFM) in tapping mode. Data analysis was performed using the freely available WSxM software \cite{Horcas-2007} and root mean square roughness (RMS) values were determined from $\unit[20\times20]{\mu m^2}$ areas. Finally, in order to directly evaluate the thermal transformation behaviour, temperature dependent electrical resistivity measurements were performed in four-contact geometry using a Quantum Design Physical Properties Measurement System (PPMS).

\section{Results}
\subsection{Room temperature deposition and post-annealing - route \textbf{A}}
\label{sec:Annealing}
Similar to already reported studies on silicon substrates \cite{Miya-2009}, our observations show that NiTi films deposited at room temperature on single crystalline MgO(001) substrates also exhibit amorphous structures. We therefore selected annealing temperatures between \unit[450]{\degree} and \unit[850]{\degree}, close to those reported in the literature. Annealing times were selected as 60 and \unit[120]{min}, respectively.

\begin{figure}[htb]
	\centering
  \includegraphics[width=0.5\textwidth]{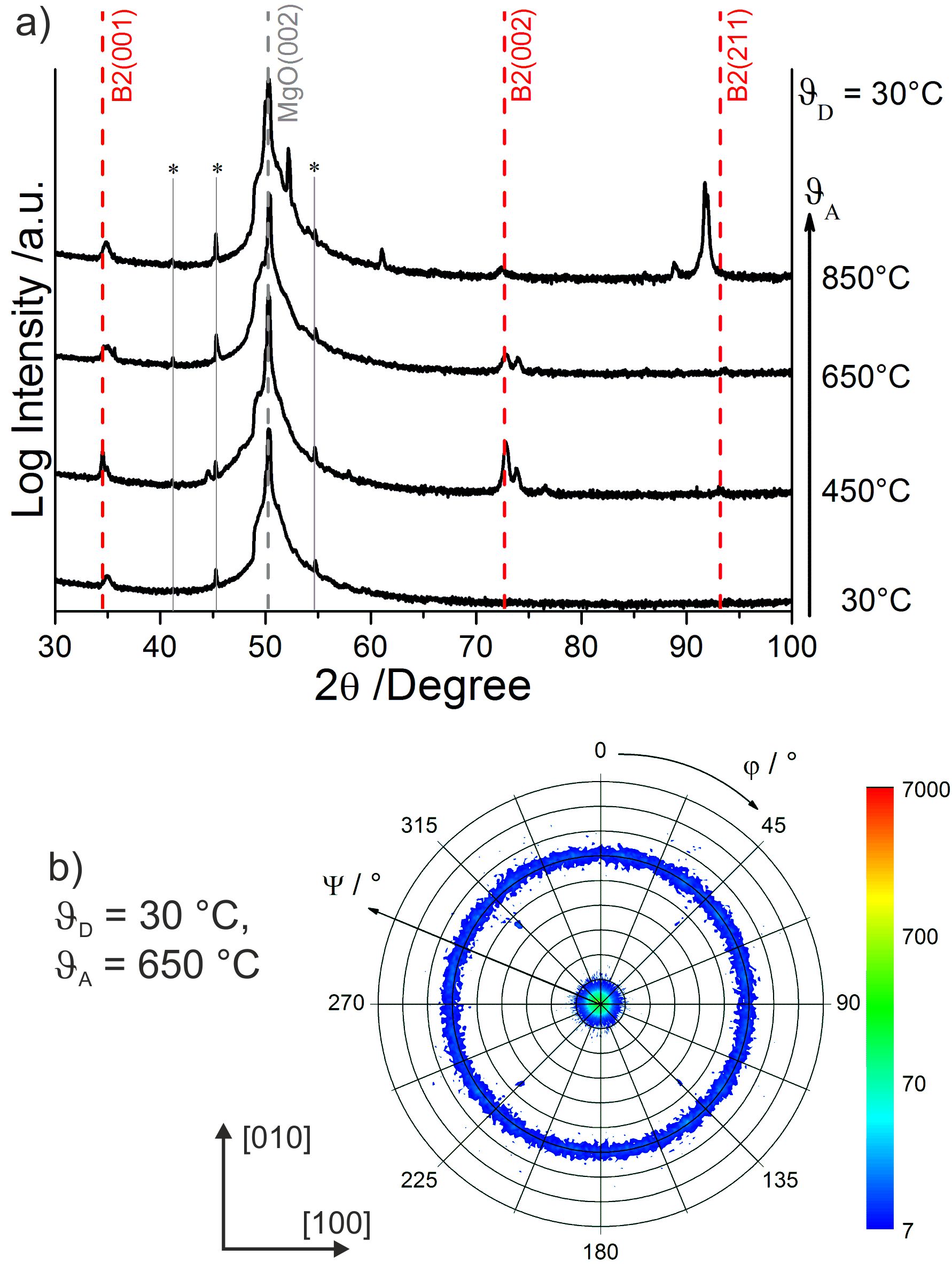}
	\caption{Route \textbf{A}: (a) Image section of the measured XRD $\theta$-2$\theta$ scans of Ni–Ti films on MgO(001) substrates deposited at $\td{}=\unit[30]{\degree}$ and annealed at different temperatures $\vartheta_{\text{A}}$. Dashed lines mark the reflection positions of bulk Ni-Ti $B2$. Reflections marks by stars arises from the XRD set-up. (b) Typical $\left\{101\right\}_{\text{B}2}$ pole figure measured of the film deposited at $\td{}=\unit[30]{\degree}$ and annealed at $\ta{}=\unit[650]{\degree}$}
	\label{fig:1}
\end{figure}

Fig.~\ref{fig:1}a shows selected XRD scans in $\theta$-2$\theta$ geometry for different annealing temperatures $\vartheta_{\text{A}}$ and an annealing time of \unit[120]{min}. For the as-deposited film ($\td=\unit[30]{\degree}$) no NiTi reflections are observed due to the amorphous structure. Annealing between \unit[450]{\degree} and \unit[650]{\degree} leads to crystallization in the B2 phase, as indicated by the $(001)_{\text{B}2}$, $(002)_{\text{B}2}$ and $(211)_{\text{B}2}$ reflections. Furthermore, a splitting of the $(001)_{\text{B}2}$ reflection is observed, which is usually discussed as the formation of \textit{R}-Phase via a martensitic transformation where the B2 austenite cell is subjected to a rhombohedral distortion \cite{Otsuka-2005}. A more detailed discussion on the observed peak splitting is given below (section \ref{sec:HeatDepo}). When increasing annealing temperature to \unit[850]{\degree}, the intensity of the $(002)_{\text{B}2}$ reflection decreases and another reflection close to $(211)_{\text{B}2}$ increases. This confirms that grains with a $(211)_{\text{B}2}$ orientation are formed. 

Since XRD scans in $\theta$-2$\theta$ geometry are only suitable to determine out-of-plane lattice parameters, pole figures of $B2$ reflections were also measured to determine the global texture of the films. For example, we show in Fig.~\ref{fig:1}b the $(101)_{\text{B}2}$ pole figure of the film deposited at room temperature and subsequently annealed at $\ta{}=\unit[650]{\degree}$ for $t_{\text{A}}=\unit[120]{min}$. Besides poles of a minor epitaxially grown NiTi represented by weak intensities at $\Psi=45^\circ$ and $\Phi=45^\circ$, a prominent ring at $\Psi=60^\circ$ is observed. This ring is caused by a pronounced $(101)_{\text{B}2}$ fibre texture. We note that the $(101)_{\text{B}2}$ reflection cannot be observed in $\theta$-2$\theta$ XRD scans because it overlaps with the high intensity (002) reflection of the MgO substrate. These complementary pole figure data confirm that the annealed films following route \textbf{A} are polycrystalline. 

\begin{figure}[htb]
	\centering
  \includegraphics[width=0.5\textwidth]{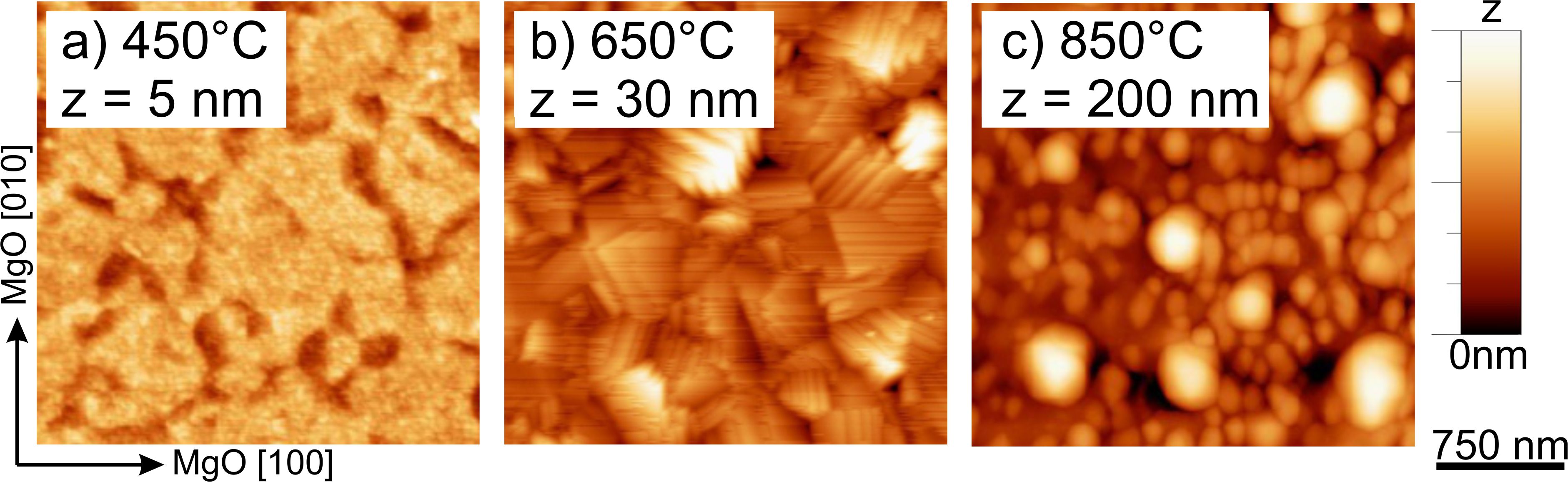}
	\caption{Route \textbf{A}: (a- c) AFM images of NiTi films annealed at \unit[450]{\degree}/\unit[120]{min}, \unit[650]{\degree}/\unit[120]{min} and \unit[850]{\degree}/\unit[60]{min}. In order to highlight the differences in morphology, the \textit{z}-scale is different for all micrographs.}
	\label{fig:2}
\end{figure}

To investigate the influence of annealing temperature on surface morphology, AFM measurements were performed. After deposition at room temperature, the film morphology is characterized by randomly arranged fused particles and by a low roughness value of RMS\,=\,\unit[0.6]{nm} (see also Fig.~\ref{fig:2}), although these films are amorphous and an additional annealing treatment is required for crystallization. Fig.~\ref{fig:2} shows AFM images for different annealing temperatures \ta. For $\ta{}=\unit[450]{\degree}$, a discontinuous, porous film is observed. The surface profile of the film annealed at $\ta{}=\unit[650]{\degree}$ shows differently arranged plates. A further increase of the annealing temperature to \unit[850]{\degree} leads to de-wetting effects and again no dense, continuous film can be obtained. 
In general, increasing the annealing temperature \ta\ results in an increased surface roughness. The maximum \textit{z}-value increases from \unit[4]{nm} for the room temperature deposited film to \unit[200]{nm} for the annealed film at \unit[850]{\degree}, whereby, for the later sample, the \textit{z}-value exceeds the average film thickness. For the sample annealed at \unit[850]{\degree} determining physically reasonable RMS values is prone to errors. These films tend to exhibit phase separation and discontinuous growth. Shorter annealing times result in less dewetting and a more continuous film.

\begin{figure}[htb]
	\centering
  \includegraphics[width=0.5\textwidth]{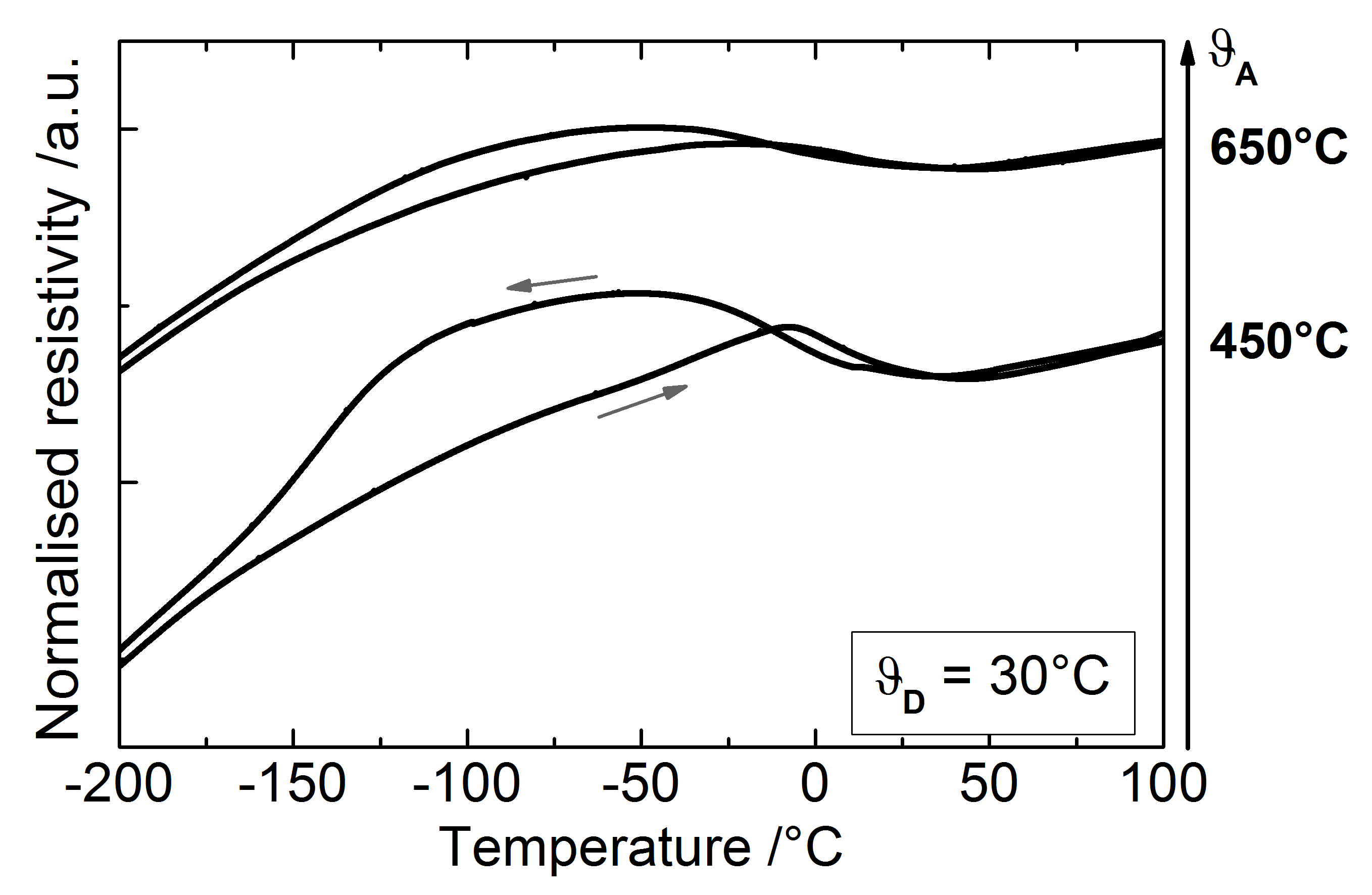}
	\caption{Route \textbf{A}: Characteristic \emph{R}(T) curves of reversible phase transformations in NiTi films of the samples shown in Fig.~\ref{fig:2}a) and b). The curves are offset for clarity and normalized to resistivity values at room temperature.}
	\label{fig:3}
\end{figure}

Temperature dependent resistivity measurements were performed in order to analyse the thermal transformation behaviour. Fig.~\ref{fig:3} shows \emph{R}(T) curves for two films. After annealing at $\ta{}=\unit[450]{\degree}$ or at $\ta{}=\unit[650]{\degree}$ ($t_{\text{A}}=\unit[120]{min}$ in both cases), the two-stage B2$\leftrightarrow$R$\leftrightarrow$B19' transformation can be observed. For all films the austenite finish temperature is close to room temperature; the R(T) data confirm that $B2$ and R-phase can co-exist in this temperature range. Compared with the hysteresis of the NiTi films deposited at higher temperatures (see next sections, e.g., Fig.~\ref{fig:6}), the area of the hysteresis is much smaller. This can be attributed to the polycrystalline microstructure of the films following route \textbf{A} compared to the epitaxial films of routes \textbf{B} and \textbf{C}. 

To sum up our results on films produced by route \textbf{A}, room temperature deposition results in amorphous films with very low roughness values. Annealing of these films, however, results in polycrystalline microstructures with only a small amount of epitaxially grown NiTi. During annealing, the formation of the B2 phase takes place. These films therefore show a reversible thermal transformation behavior as expected for polycrystalline NiTi thin film samples. The overall roughness of the films increases with increasing annealing temperature. 

\subsection{Deposition on heated substrates - route \textbf{B}}
\label{sec:HeatDepo}
Analysis of films deposited at room temperature clearly showed that crystallization of B2 NiTi occurs during post-deposited heating of the samples. We therefore devised preparation route \textbf{B}, where a higher mobility of atoms required for crystallization are already provided during deposition via  heating of the substrate. In contrast to deposition at room temperature with unheated substrates in route \textbf{A}, the NiTi thin films deposited on the heated substrates exhibit crystalline structures even without an additional annealing treatment. 

\begin{figure}[htb]
	\centering
  \includegraphics[width=0.5\textwidth]{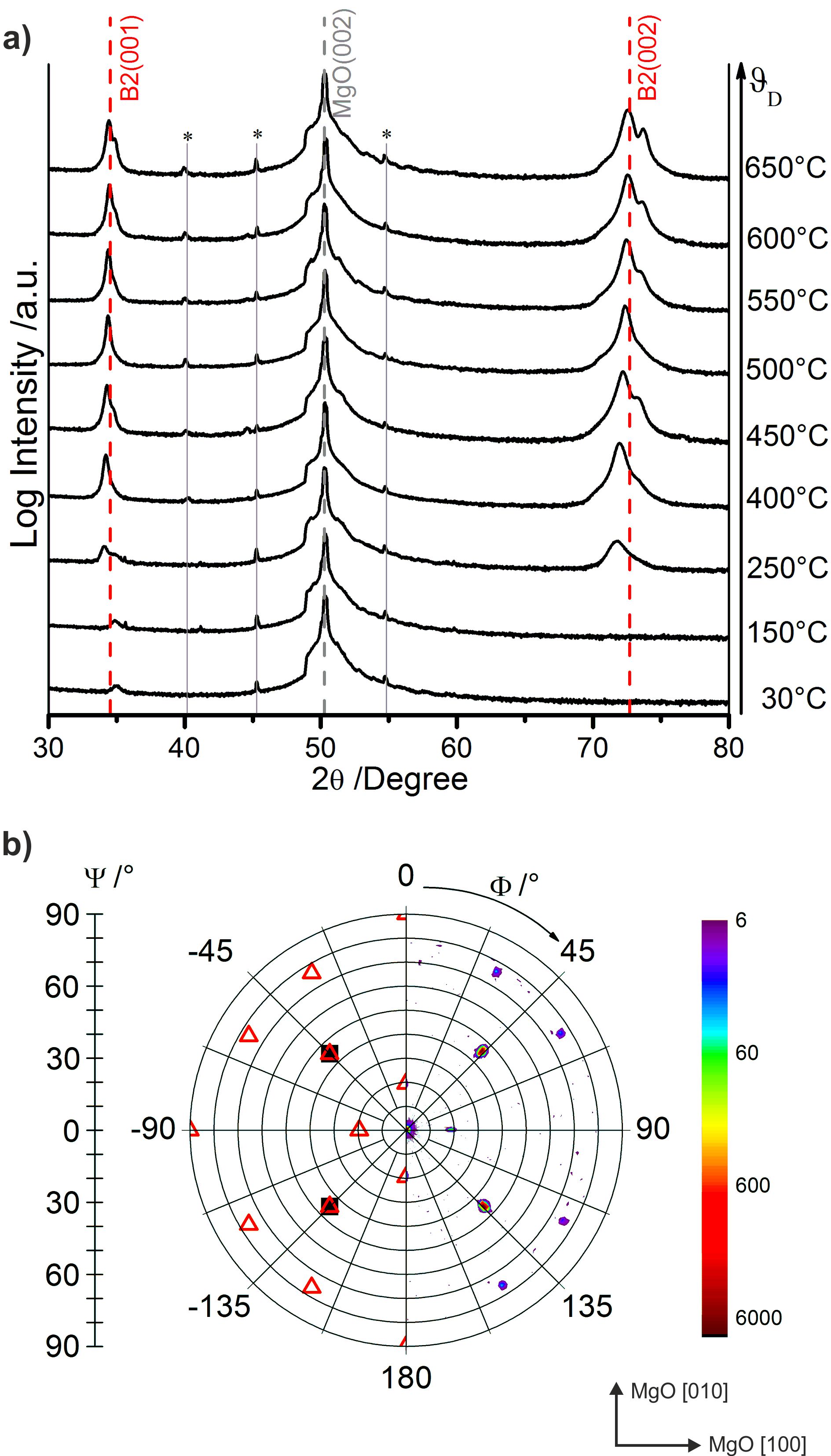}
	\caption{Route \textbf{B}: a) Image section of the measured XRD $\theta$-2$\theta$ scans of NiTi films on MgO(001) substrates grown at different deposition temperatures $\vartheta_{\text{D}}$. Dashed lines mark the reflection positions of bulk NiTi $B2$. Reflections marked by stars result from the XRD set-up. b) The left side of the depicted pole figure is a calculated one with a simulation of $\left\{112\right\}_{\text{bcc}}\left\langle 11\bar{1}\right\rangle_{\text{bcc}}$ twinning (left side). Black squares represent the epitaxial fraction of the film and red triangle the \emph{bcc} growth twins. The right side of the pole figure shows the corresponding measured data.}
	\label{fig:4}
\end{figure}

XRD scans for different deposition temperatures $\vartheta_{\text{D}}$ are shown in Fig.~\ref{fig:4}a. For films with $\td\leq\unit[150]{\degree}$, no reflections originating from the NiTi film occur. With increasing deposition temperature, all films exhibit high intensity reflections close to positions of the  $(001)_{\text{B}2}$ and $(002)_{\text{B}2}$ reflections of the NiTi $B2$ phase. Increasing the deposition temperature \td{} also results in a shift of the $(00l)_{B2}$ reflection to higher $2\theta$-values. Similar to the observations on the annealed samples of route \textbf{A}, double reflections close to the position of $(00l)_{\text{B}2}$, which indicate lattice distortion as a result of martensitic transformation, are observed for deposition temperatures above $\td>\unit[400]{\degree}$.

In order to make sure that no other crystal orientations than $(00l)_{\text{B}2}$ are present in the samples, global texture measurements need to  be analyzed. Fig.~\ref{fig:4}b shows a $\left\{101\right\}_{\text{B}2}$ pole figure of a typical sample deposited at $\td=\unit[450]{\degree}$. For this measurement, the MgO[100] edges are oriented parallel to the edges of the figure. The four well-defined sharp peaks with high intensities indicate a high-quality epitaxial growth of NiTi. The $\left\{101\right\}_{\text{B}2}$ pole is rotated by $45^\circ$ with respect to the substrate edges of the MgO substrate. A sample tilt angle of $\Psi=\,45^\circ$ and rotation angles of $\Phi=\,n\cdot 45^\circ$ are observed. The pole at $\Psi=\,0^\circ$ and $\Phi=\,0^\circ$ originates from the $\left\{002\right\}_{\text{MgO}}$ reflections of the substrate. Clearly, the cubic unit cells of the thin NiTi film and the MgO substrate are rotated by $45^\circ$ with respect to each other, which corresponds to the orientation relationship MgO(001)[100]$||$NiTi$_{\text{B2}}$(001)[110]. In contrast to results from Buschbeck et al.~\cite{Buschbeck2011} on very thin NiTi films on MgO(001) deposited by molecular beam epitaxy, we do not observe a substantial tetragonal distortion of the $B2$ unit cell that would lead to a shift of the 101 pole in $\Psi$-direction.

The origin of the peak shift in $2\theta$ XRD measurements considering the splitting of the B2(00l)-reflection at higher deposition temperatures can be explained as follows: lattice parameters (\emph{a}) and thermal expansion coefficient $\alpha$ of the MgO substrate and the NiTi film are different. As a simple means to illustrate the difference, we consider the lattice mismatch between two distinguished lattice planes. With the observed epitaxial orientation relationship, the lattice mismatch is calculated with $m=(\sqrt{2}a_{\text{B2}}-a_{\text{MgO}})/a_{\text{MgO}}$. At room temperature ($a_{\text{NiTi}}=\unit[0.3015]{nm}$, $a_{\text{MgO}}=\unit[0.4212]{nm}$), the lattice mismatch is $m =\unit[-1.3]{\%}$, which leads to compressive stresses within the NiTi film plane. With increasing temperature, the lattice parameter of NiTi increases more strongly than the lattice parameter of MgO, and thus cooling from deposition temperatures to room temperature also results in compressive stresses.

In addition to the high intensity signals at $\Psi=45^\circ$ and $\Phi=n\cdot 45^\circ$ marked by black squares, which originate from the epitaxially grown NiTi, further signals with intensities lower than $\unit[1]{\%}$ of the main signals are observed, which become visible only in the logarithmic scale used. To further understand the origin of these lower intensity signals, simulations of pole figures were carried out, following the procedure outlined in \cite{Sandra-Metals}. Using the substrate as an absolute reference frame, the pole figure simulations predict signals as depicted in the left quadrant in Fig.~\ref{fig:4}b as triangles. Compared to the measured signals the simulations may identify  $\left\{112\right\}_{\text{bcc}}\left\langle 11\bar{1}\right\rangle_{\text{bcc}}$ twinning of the $B2$ austenite as origin of the low intensity signals. This  $\left\{112\right\}_{\text{bcc}}\left\langle 11\bar{1}\right\rangle_{\text{bcc}}$ twinning is a common twinning mode in \emph{bcc} metals \cite{Christian-1995}. Twinning can occur during different processes, like deformation, film growth, recrystallisation, phase transformation, or adaptive nanotwinning. The most likely mechanism that can account for the observed twin formation in our NiTi films is the strained film growth itself. We note that the well-defined twinning relation does not give any additional reflections perpendicular to the substrate and therefore $\left\{112\right\}_{\text{bcc}}\left\langle 11\bar{1}\right\rangle_{\text{bcc}}$ twinning cannot be identified in $\theta$-2$\theta$ geometry. Considering the different heat treatments studied here, decreasing deposition temperatures lead to higher amounts of $\left\{112\right\}_{\text{bcc}}\left\langle 11\bar{1}\right\rangle_{\text{bcc}}$ twins, whereas increasing deposition temperatures suppresses the formation of twins. We attribute this observation to an increased relaxation by misfit dislocations which is a thermally activated process.

\begin{figure*}[htb]
	\centering
  \includegraphics[width=0.9\textwidth]{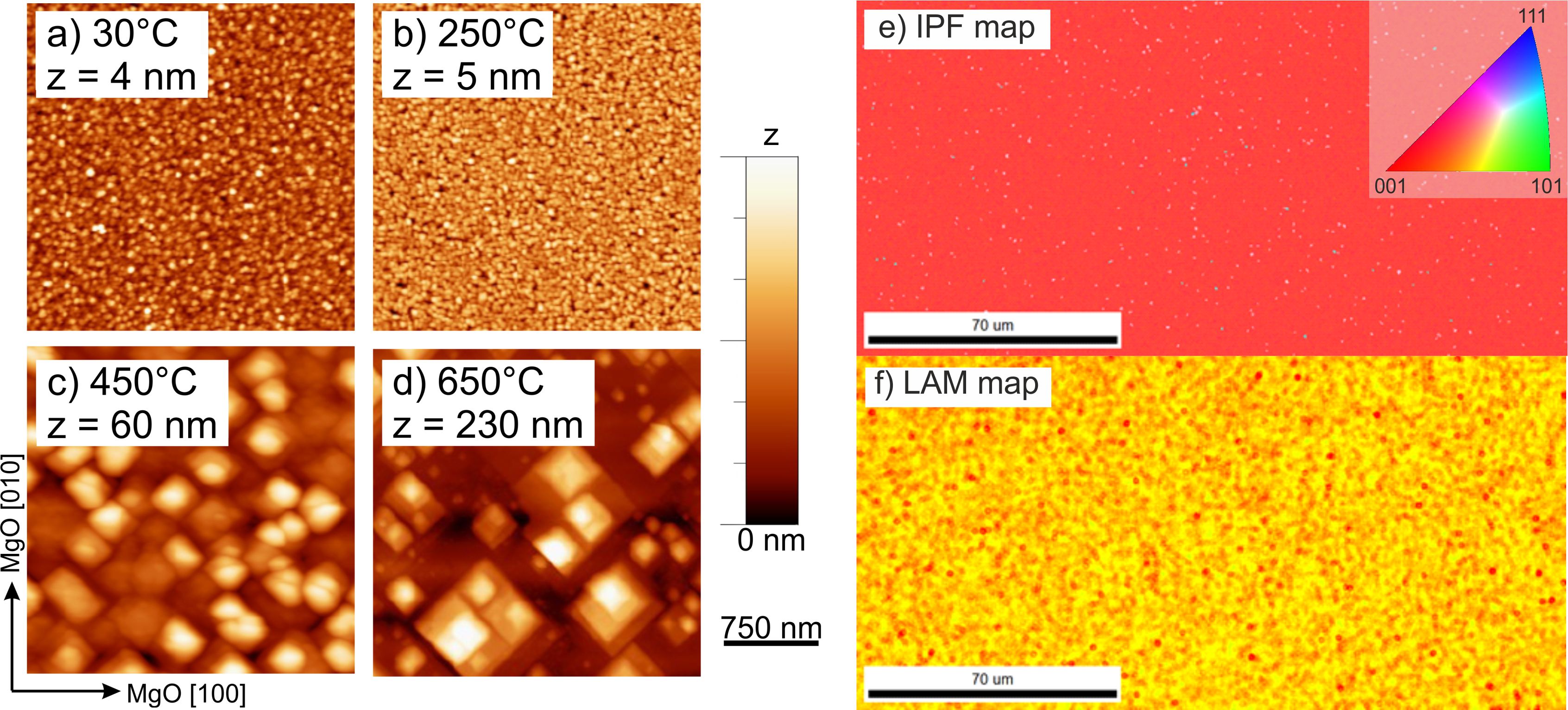}
	\caption{Route \textbf{B}: a)-d) AFM images of films with different deposition temperatures $\vartheta_{\text{D}}$. In order to highlight the differences in morphology, the \emph{z}-scale is different for all micrographs. e) Inverse pole figure (IPF) map from EBSD measurements of the normal film plane scanned over a large sample area. The red color indicates a well-defined (001) orientation of the sample deposited at \unit[550]{\degree}. f) Local average misorientation (LAM) map from EBSD of the sample shown in c). The yellow color corresponds to misorientations up to $\unit[1]{^\circ}$, the red color indicates misorientations between $\unit[1]{^\circ}$ and $\unit[5]{^\circ}$.}
	\label{fig:5}
\end{figure*}

In Fig.~\ref{fig:5}a - d, AFM images of selected films with characteristic features are presented. As a reference, an AFM scan of a film deposited at room temperature (characterized by the lowest measured roughness value of \unit[0.6]{nm}) is shown in Fig.~\ref{fig:5}a. As observed on samples prepared by route \textbf{A}, film roughness generally increases with increasing deposition temperatures \td{}. In Fig.~\ref{fig:5}b no regularly patterned surface is observed; the roughness value is slightly increased to RMS\,=\,\unit[0.7]{nm}. For $\td\geq\unit[450]{\degree}$ (Fig.~\ref{fig:5}c), a regularly patterned surface exhibiting square features occurs. The squares are rotated by $45^\circ$ with respect to the MgO[100] edges and become larger with increasing deposition temperature, see Fig.~\ref{fig:5}d. At $\td{}=\unit[650]{\degree}$ the RMS roughness is \unit[48]{nm}, which is a substantial fraction of the experimental film thickness of \unit[120]{nm}. The orientation of the squares is similar to the orientation of the NiTi $B2$ unit cell determined by pole figure measurements. This observation suggests that $\left\{001\right\}$ surfaces have the lowest surface energy of the ordered $B2$ phase. Moreover, increasing roughness and formation of growth twins exhibit the opposite temperature dependency. As mentioned above, an increasing deposition temperature suppresses the formation of $\left\{112\right\}_{\text{bcc}}\left\langle 11\bar{1}\right\rangle_{\text{bcc}}$ twins; simultaneously, the roughness increases and a regularly patterned surface is formed.

In very thin NiTi films, mismatch stresses can result in a tetragonal distortion, as observed by Buschbeck et al.~\cite{Buschbeck2011} in epitaxial NiTi films with a thickness of \unit[35]{nm}. In thicker epitaxial films, stresses can be reduced by several processes: One possibility is the installation of misfit dislocations which become more mobile at higher temperatures. This effect may be indicated by the rough, patterned surface at higher deposition temperature of our route \textbf{B} (see Fig.~\ref{fig:5}d). The changes in growth mode from a smoothed layer to layer mode to a patterned surface appear for a deposition temperature above $\td\geq\unit[250]{\degree}$. At this temperature, NiTi is grown mostly epitaxially, and the roughness increases. As described in \cite{Harsha2006}, a competition between elastic energy and surface energy defines this instability. Surfaces under compressive stress become rougher. The equilibrium shape in many materials shows faceting and surface steps. A second possibility for reducing mismatch stresses is the formation of growth and/ or deformation twins -- as discussed above, this may well be the case in our samples with a deposition temperature about $\td=\unit[450]{\degree}$ (see Fig.~\ref{fig:4}b). A third possibility is a stress induced (pre-)martensitic transition to the \emph{R}-phase, as indicated by the splitting of the B2(00l)-reflection for deposition temperatures above the austenite finish temperature \cite{Otsuka-2005}.

Local texture measurements with EBSD (Fig.~\ref{fig:5}e and f) confirm that the film grows epitaxially in $B2$ phase. The inverse pole figure (IPF) map shown in Fig.~\ref{fig:5}e also confirms the well oriented film growth. The different shades of red indicate deviations from the Kikuchi pattern of a perfectly aligned austenitic unit cell up to $\unit[5]{^\circ}$ in each measured spot. This is visible more clearly in the local average misorientation (LAM) map, Fig.~\ref{fig:5}f. The third order averaged misorientation was determined as approximately $\unit[0.4]{\%}$, which again confirms the high quality of the epitaxial growth.

\begin{figure}[htb]
	\centering
  \includegraphics[width=0.5\textwidth]{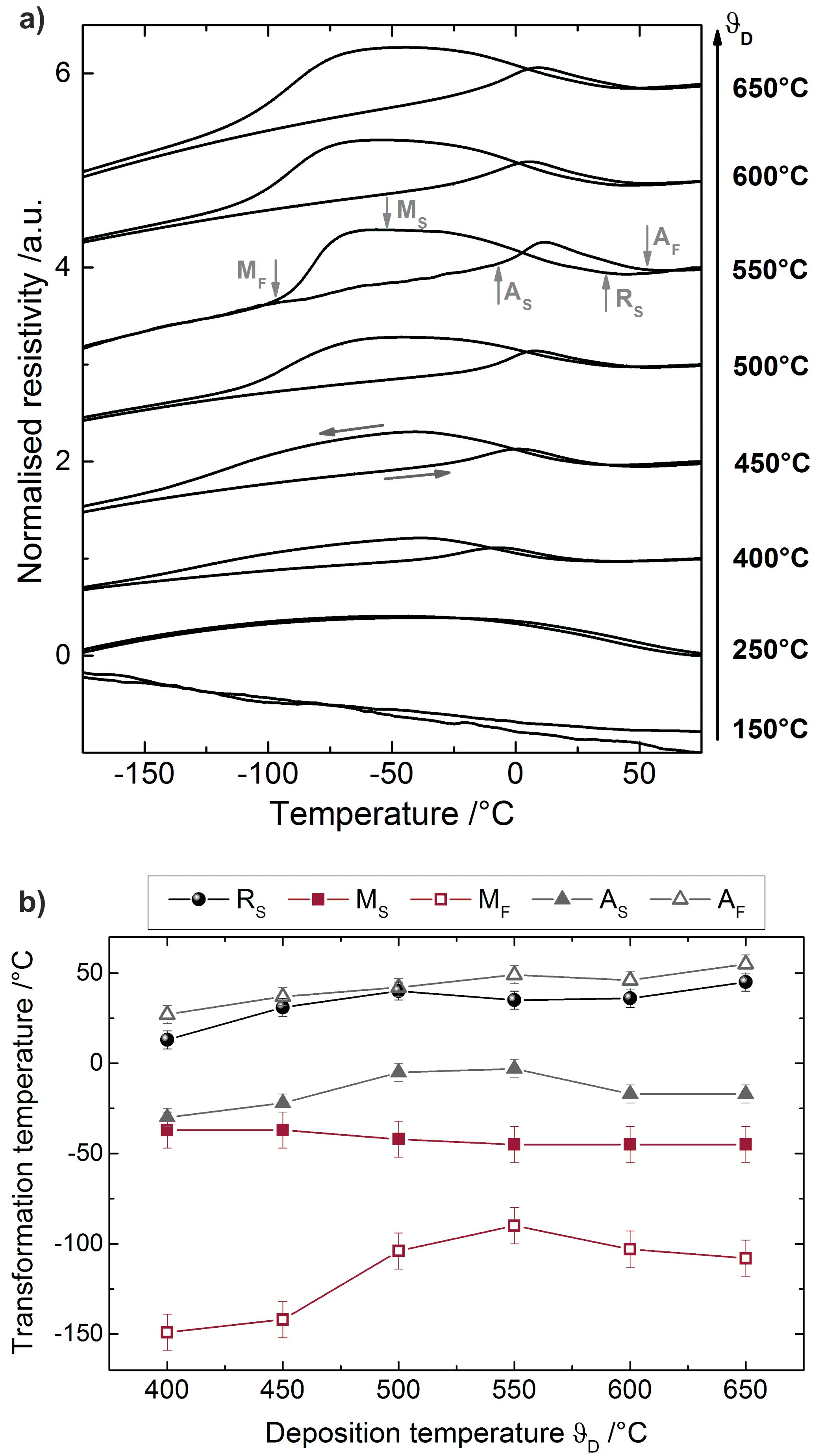}
	\caption{Route \textbf{B}: a) Characteristic \emph{R}(T) curves of reversible phase transformations of $B2\leftrightarrow R\leftrightarrow B19'$ and $B2\leftrightarrow R$ in films prepared with different deposition temperatures $\vartheta_{\text{D}}$. The curves are offset for clarity and normalized to room temperature. b) Transformation temperatures obtained from resistivity measurements in a) as a function of deposition temperature.}
	\label{fig:6}
\end{figure}

The results discussed so far demonstrate that thin NiTi films deposited at sufficiently high temperatures in principle exhibit the required microstructural features that are needed if one is interested in using the thin films as model systems to further study martensitic transformations in NiTi. We now present the results of electrical resistivity measurements to prove that the thin films actually exhibit martensitic transformation(s) during cooling. Fig.~\ref{fig:6}a) shows characteristic temperature dependent resistivity-temperature \emph{R}(T) curves for different deposition temperatures. With respect to the phase assignment we follow the well-established interpretation of the \emph{R}(T) curves, as given e.g. by Otsuka and Ren \cite{Otsuka-2005}. Fig.~\ref{fig:6}a) shows that the deposition temperature has an influence on the curve shape that represents different reversible phase transformations. For deposition temperatures below \unit[250]{\degree}, resistivity continuously increases with decreasing temperature. This so-called semiconductor behavior is often attributed to the formation of strain glass in alloys with quenched disorder.\cite{Ren-2011}

In films deposited at higher temperatures, reversible martensitic transformations are observed. For $\td{}=\unit[400]{\degree}$, a $B2\leftrightarrow R$ phase transformation with a very small hysteresis occurs. For higher deposition temperatures, films exhibit a two-stage $B2\leftrightarrow R\leftrightarrow B19'$ transformation: During cooling, a deviation from linearity and an increase in resistivity marks the start of the \emph{R}-phase formation. The maximum in resistivity is observed when the inter-martensitic transformation starts and the \emph{R}-phase as well as the monoclinic martensitic $B19'$ phase coexist. This transformation is completed once the resistivity continuously decreases linearly. Upon heating, the $B19'$ martensite is transformed directly to austenite and a weak increase in resistivity is observed in the temperature range where both structures coexist.

The characteristic phase transformation temperatures were determined from the \emph{R}(T) curves and are summarized in Fig.~\ref{fig:6}b). With increasing deposition temperature the transformation temperatures increase. For deposition temperatures above $\unit[500]{\degree}$ the \emph{R}-phase start temperature ($R_\text{S}$) is slightly above room temperature. The correlation between $R_\text{S}>\unit[30]{\degree}$ from our electrical measurements and the splitting of X-ray reflections (Fig.~\ref{fig:4}a)) clearly show that $B2$ austenite and \emph{R}-phase may coexist at room temperature. The small distortions of the $B2$ crystal lattice associated with formation of the \emph{R}-phase are hardly detectable by EBSD: They are only indicated as very small changes in confidence index values. This explains why our EBSD measurements only reveal the cubic $B2$-phase. Furthermore, the martensite finish temperature of monoclinic martensite $B19'$ ($M_\text{F}$) also increases with increasing deposition temperature. For $\td\geq\unit[500]{\degree}$ the martensitic transformation is completed at \unit[-100]{\degree}. The effect of deposition temperature on $M_\text{F}$ is considerably more pronounced than on $R_\text{S}$. Compared to the transformation temperatures of very thin epitaxial films~\cite{Buschbeck2011}, polycrystalline films and bulk materials~\cite{Miyazaki1999,Frenzel2010} with a Ni$_{50}$Ti$_{50}$ composition, the transformation temperatures of our epitaxial films are slightly higher. This is most likely related to the high stresses as described before.

\subsection{Deposition at \unit[250]{\degree} and post-annealing - route \textbf{C}}
\label{sec:warm}
\begin{figure*}[htb]
	\centering
  \includegraphics[width=0.7\textwidth]{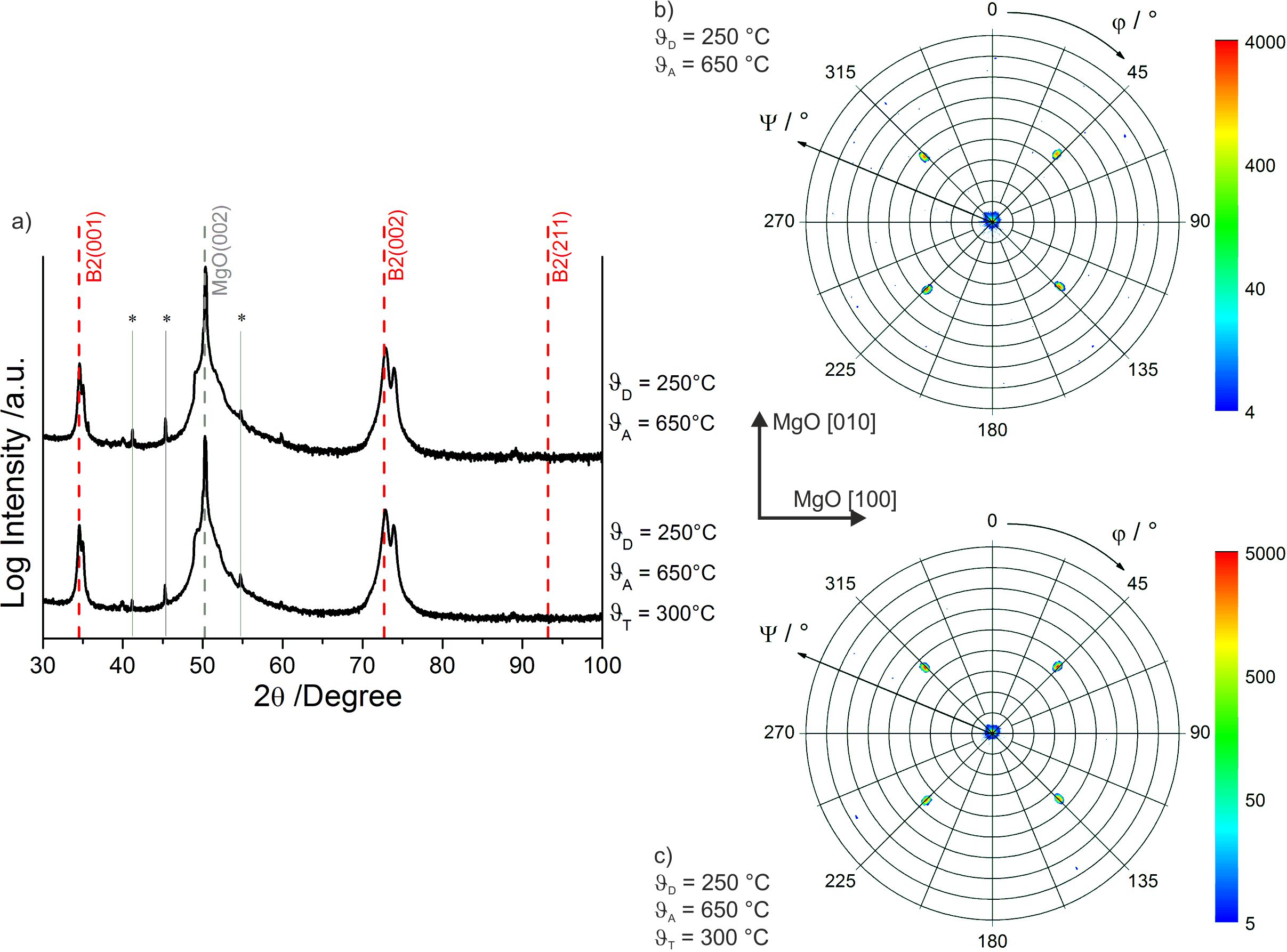}
	\caption{Route \textbf{C}: a) Image section of the XRD $\theta$-2$\theta$ scans of NiTi films deposited at $\ta{}=\unit[250]{\degree}$ followed by annealing at $\ta{}=\unit[650]{\degree}$ and an additional tempering at $\temp{}=\unit[300]{\degree}$ for \unit[30]{min}. Dashed lines mark the reflection positions of bulk NiTi $B2$. Reflections marked by stars result from the XRD set-up. b) $\left\{101\right\}_{\text{B}2}$ pole figure of a film deposited at $\td{}=\unit[250]{\degree}$ and annealed at $\ta{}=\unit[650]{\degree}$ and c) with an additionally tempered at $\temp{}=\unit[300]{\degree}$.}
	\label{fig:7}
\end{figure*}

In general, the deposition conditions applied within route \textbf{B} result in epitaxial microstructures that exhibit a reversible thermal martensitic transformation. However, the relatively high surface roughness of the films produced by route \textbf{B} are detrimental, for instance, for further mechanical characterization by nanoindentation. We therefore investigated whether deposition at moderate temperatures followed by a post-deposition annealing and an additional tempering step (\unit[300]{\degree}) could result in films that are both epitaxial and smooth. This route \textbf{C} should combine advantages of both previous processes \textbf{A} and \textbf{B}. As evident from the XRD measurements in Fig.~\ref{fig:7}a), deposition at $\td{}=\unit[250]{\degree}$ is sufficient for the mostly epitaxial crystallization of the $B2$ phase. We therefore selected this temperature for deposition of all samples considered following route \textbf{C}. However, without an additional heat-treatment, these films do not transform, indicating an incomplete order and the presence some bcc growth twins. Therefore, films were post-deposition annealed at \unit[650]{\degree}. This temperature is commonly used for solution annealing in bulk NiTi~\cite{Miya-2009}. Furthermore, this annealing temperature leads to transforming films in route \textbf{A}. This, however, results in an increase of surface roughness. Therefore, after the post-deposition annealing an additional temper step at \unit[300]{\degree} was carried out for \unit[30]{min}. In contrast to the high temperature annealing steps, no solution takes place during tempering, and the enhanced diffusion allows for surface rearrangement and therefore promotes the formation of smoother films. The main results of these experiments are summarized below. 

\begin{figure*}[htb]
	\centering
  \includegraphics[width=0.9\textwidth]{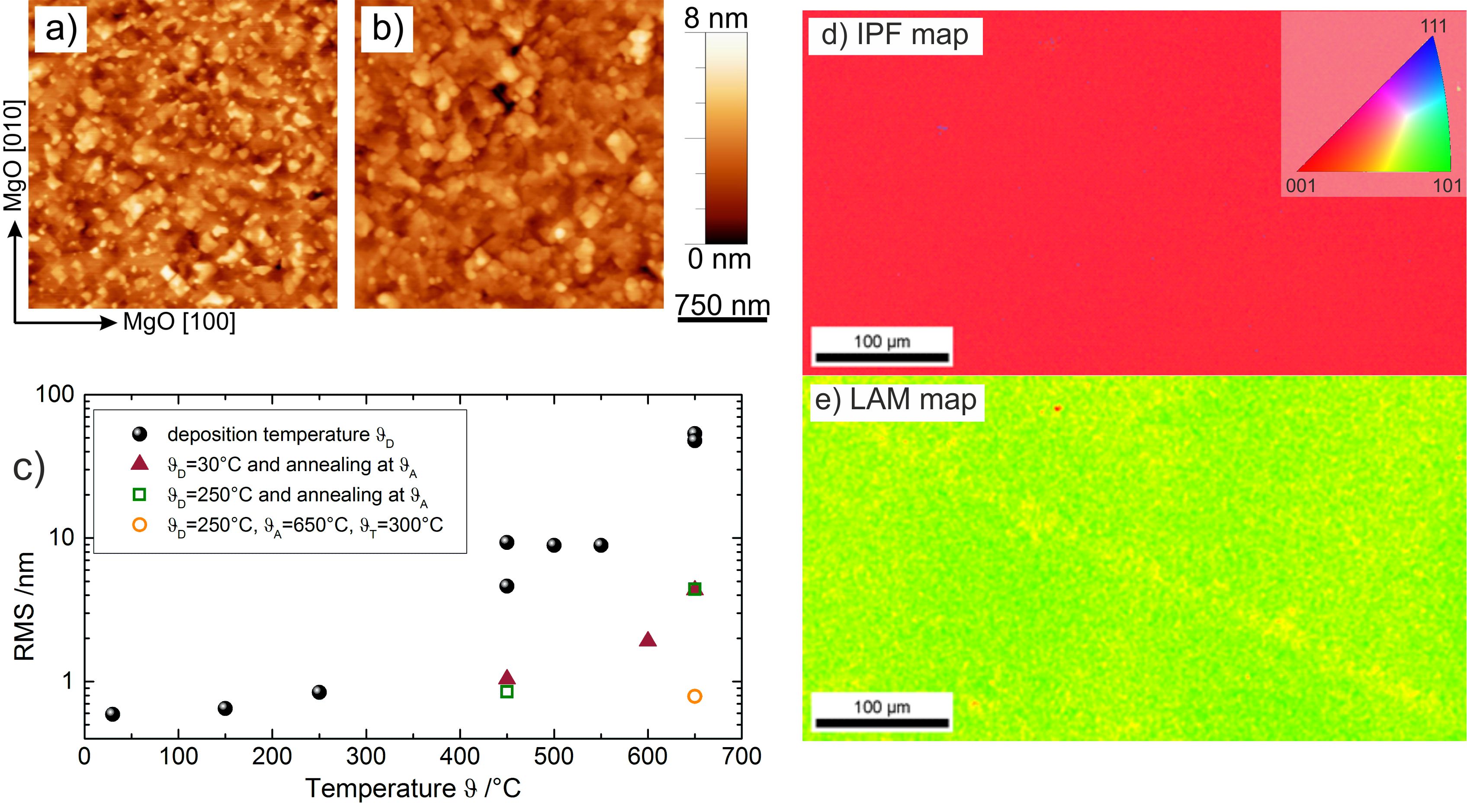}
	\caption{Route \textbf{C}: a) AFM images of annealed NiTi films ($\vartheta_{\text{D}}=\unit[250]{\degree}$, $\vartheta_{\text{A}}=\unit[650]{\degree}$) and b) additional tempering with $\temp{}=\unit[300]{\degree}$. c) Summary of the influence of deposition temperature \td\ and/or annealing temperature \ta\ on RMS values of \unit[120]{nm} thick NiTi films. d) Inverse pole figure map of the normal film plane and e) local average misorientation map from the EBSD measurements of the tempered film. The green colour corresponds to misorientations smaller than $\unit[1]{\degree}$.}
	\label{fig:8}
\end{figure*}

For reference, XRD scans of the annealed -- but not tempered -- NiTi films deposited at $\td{}=\unit[30]{\degree}$ and $\td{}=\unit[250]{\degree}$ are shown in Fig.~\ref{fig:7}a). Both annealed films are characterized by the splitting of the $(00l)_{\text{B}2}$ peak already discussed above. The normalized intensity for annealed films with $\td{}=\unit[30]{\degree}$ from route \textbf{A} is lower than for annealed films with $\td{}=\unit[250]{\degree}$ of route \textbf{B}. This can be explained by the polycrystalline microstructure associated with a fibre texture microstructure of the route \textbf{A} film which consists of very small grains in the range of a several nanometres. In contrast, the film prepared by route \textbf{B} is already epitaxially grown during deposition. Deposition at $\td{}=\unit[250]{\degree}$ leads to crystallization in the $B2$ phase, further growth during the annealing step, and results in a well oriented, basically single crystalline structure, see global pole figure measurements in Fig.~\ref{fig:7}b). AFM data presented in Fig.~\ref{fig:8}a) shows a relatively smooth surface and results in a RMS roughness of \unit[5]{nm}.

The additional temper step at \unit[300]{\degree} represents a potential approach to reduce the roughness of the NiTi films: An XRD scan of a typical tempered sample is shown in Fig.~\ref{fig:7}a. Compared to the high temperature annealed films, the intensity of the split B2(00l) reflection is similar to the one of the sample prepared by route \textbf{B}. This indicates that the NiTi films maintain their epitaxial features during route \textbf{C} preparation. Global texture measurements also ascertain the epitaxial growth of the film, Fig.~\ref{fig:7}c, as do the local texture measurements with EBSD (Fig.~\ref{fig:8}d and e). Compared to the EBSD results for route \textbf{B}, lower variations in the Kikuchi patterns of an austenitic unit cell are observed. The third order averaged misorientation shown in Fig.~\ref{fig:8}e is much smaller than for the samples of route \textbf{B}. Furthermore, the AFM data presented in Fig.~\ref{fig:8}c shows a very smooth surface with a RMS value of about \unit[0.8]{nm}, which is substantially reduced compared to high temperature annealed film.

A summary of the roughness values obtained from all preparation routes is shown in Fig.~\ref{fig:8}c. Lowest roughness values below \unit[1]{nm} are obtained from the experiments with low deposition temperatures in route \textbf{B}. However, these films do not transform. For deposition temperatures above $\vartheta_{\text{D}}=\unit[250]{\degree}$ a pronounced two-stage B2$\leftrightarrow$R$\leftrightarrow$B19' transformation occurs, but, roughness increases by about two orders of magnitude. High temperature annealing of room temperature deposited films (route \textbf{A}) leads to intermediate roughness values up to \unit[10]{nm}, which, however, are polycrystalline. Subsequent tempering of the samples in route \textbf{C} results in very smooth surfaces with roughness values below \unit[1]{nm}.

\begin{figure}[htb]
	\centering
  \includegraphics[width=0.5\textwidth]{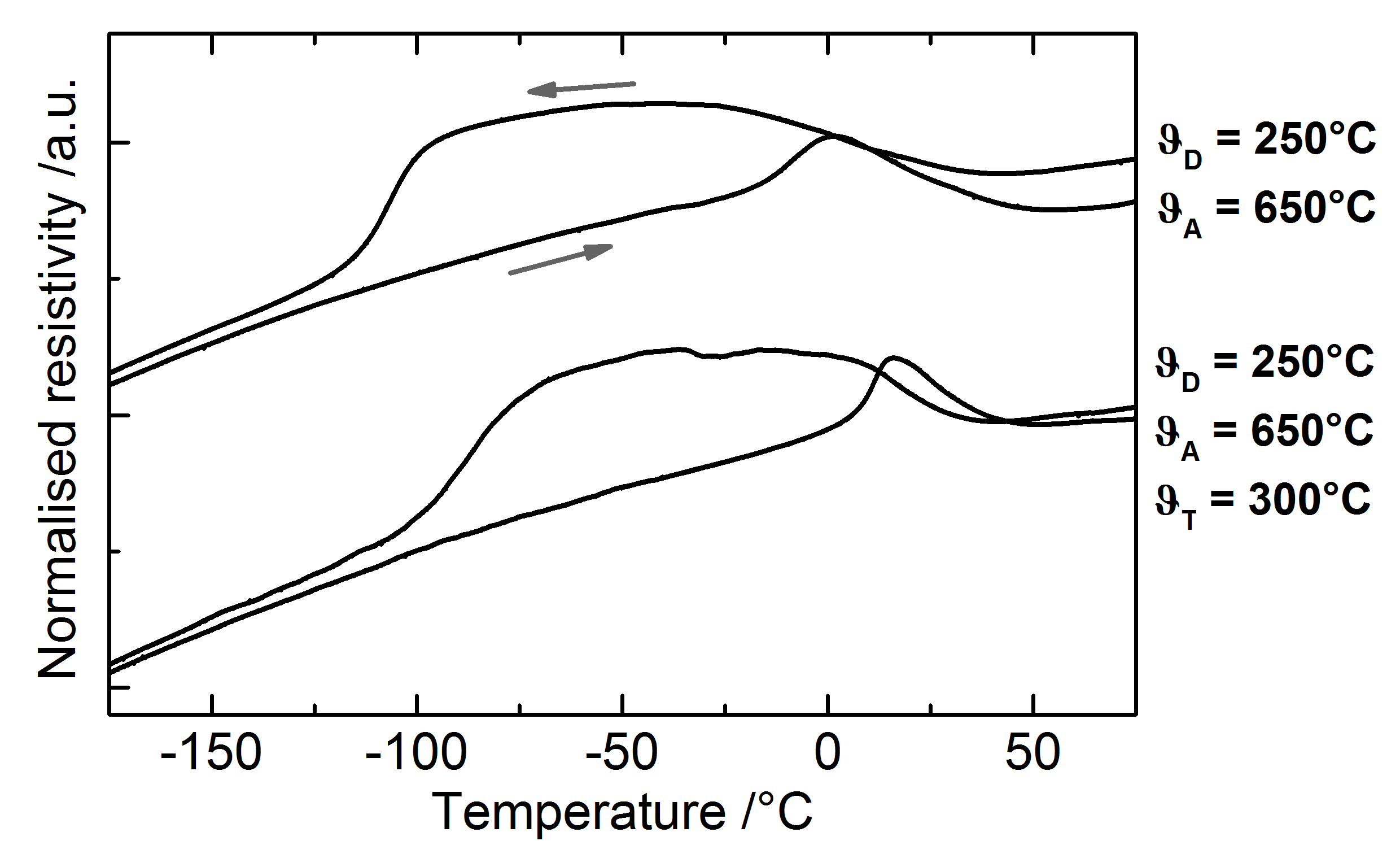}
	\caption{Route \textbf{C}: Characteristic \emph{R}(T) curves of reversible phase transformations of high temperature annealed NiTi films ($\ta{}=\unit[650]{\degree}$) and additional tempering at $\temp{}=\unit[300]{\degree}$). For clarity, the curves are offset and normalized to room temperature.}
	\label{fig:9}
\end{figure}

Finally, the tempered samples were also investigated by temperature dependent resistivity measurements. Fig.~\ref{fig:9} illustrates the corresponding \emph{R}(T) curves. The films also show a pronounced two-stage B2$\leftrightarrow$R$\leftrightarrow$B19' transformation and a larger hysteresis compared to the films prepared by route \textbf{A} and \textbf{B}. 

\section{Summary and conclusions}
\label{sec:conclusions}
In this study we investigated the epitaxial growth of NiTi films processed by DC magnetron sputtering on single crystalline MgO substrates using three different preparation routes: \textbf{A} depositing at room temperature followed by annealing, \textbf{B} depositing at high temperatures and \textbf{C} depositing at a moderately elevated temperature followed by annealing and temper steps. Our experiments demonstrate for the first time that the epitaxially grown films combine a well-defined orientation of all austenitic unit cells with smooth surfaces. Both criteria were best fulfilled by preparation route \textbf{C}. The epitaxially grown thin  films discussed in this study can be used as model system for future studies on the thermo-mechanical transformation behaviour of binary NiTi. 

Our experiments demonstrate that, in particular for the NiTi system, XRD measurements in Bragg-Brentano geometry are not sufficient to furnish proof for epitaxial growth. Neither \emph{bcc} twinning (as observed in route \textbf{B}) nor the formation of a strong $\left(101\right)$ fibre texture (as observed in route \textbf{A}) result in significant intensities in addition to $\left(00l\right)$ reflections. Therefore pole figure measurements are essential to accurately analyse orientation and microstructures of epitaxial NiTi films. 

Finally, for a qualitative understanding of the differences in topography and crystal orientation between the three routes considered in this study, two key differences in kinetics and thermodynamics need to be considered. First, surface diffusion in thin films is significantly faster than bulk diffusion because no formation of vacancies is required at a film surface. Second, surface energies determine film morphology. Smooth, continuous films only represent a metastable minimum in terms of total surface energy, whereas the global minimum is a granular film consisting of large spherical particles (Reighley instability). When films are grown at room temperature (route \textbf{A}), mobility is low and the atoms remain close to their initial positions after deposition. This results in continuous films without any tendency of dewetting. As a consequence one can observe smooth film surfaces with low roughness values. Deposition on heated substrates (route \textbf{B}) results in a sufficient surface diffusion. This allows the deposited atoms to find energetically favoured positions on the surface, which can reproduce the surface properties of the substrate. Accordingly the films can grow epitaxially at higher deposition temperatures. Deposition at moderate temperatures followed by annealing and tempering (route \textbf{C}) combines the advantages of both previously discussed routes. While the rather low deposition temperature of \unit[250]{\degree} is sufficient to promote crystallization in the $B2$ phase, and an oriented growth, the additional temper step provides enough thermal energy to increase chemical order and to remove defects like bcc growth twins. However, mobility is not high enough to leave the metastable minimum of a continuous film, which avoids dewetting.

\section{Acknowledgment}
\label{sec:acknowledgment}
The authors thank T. Grundmann and J. Scheiter for experimental support and A. Ludwig (Ruhr-Uni Bochum) for providing the Ni-Ti target.

\section*{References}
\bibliographystyle{elsarticle-num}								
\bibliography{NiTi-REFs}  													

\end{document}